\documentclass[preprint,review,1p,times]{elsarticle}
\usepackage{gensymb}
\usepackage{xcolor}
\usepackage{comment}
\usepackage{xurl}
\usepackage{float}
\usepackage{placeins}
\usepackage{lineno}

\begin{document}
\begin{frontmatter}

\title{Migration of Belle~II TOP Feature Extraction from the Zynq Processing System to PCIe40 Readout PCs}

\author[uhm]{Harsh Purwar\fnref{presentjlab}}
\author[uhm]{Shahab Kohani}
\author[uhm]{Vasily Shebalin}
\author[uhm]{Martin Bessner}
\author[uhm]{Matt Andrew}
\author[uhm]{Thomas Browder}
\author[uhm]{Tommy Lam}
\author[uhm]{Kurtis Nishimura}
\author[uc]{Sourav Patra}
\author[pitt]{Vladimir Savinov}
\author[slovenia]{Marko Stari\v{c}}
\author[uhm]{Gary Varner}
\author[kek]{Satoru Yamada}
\author[uhm]{Keisuke Yoshihara\corref{cor1}}
\ead{kyoshiha@hawaii.edu}

\cortext[cor1]{Corresponding author.}

\address[uhm]{Department of Physics and Astronomy, University of Hawaii at Manoa, Honolulu, Hawaii 96822, USA}
\address[uc]{Department of Physics, University of Cincinnati, Cincinnati, Ohio 45221, USA}
\address[pitt]{Department of Physics and Astronomy, University of Pittsburgh, Pittsburgh, Pennsylvania 15260, USA}
\address[kek]{High Energy Accelerator Research Organization (KEK), Tsukuba, Ibaraki 305-0801, Japan}
\address[slovenia]{Jo\v{z}ef Stefan Institute, Jamova cesta 39, 1000 Ljubljana, Slovenia}
\fntext[presentjlab]{Present address: Thomas Jefferson National Accelerator Facility, Newport News, Virginia 23606, USA.}

\begin{abstract}
The Belle~II Time-of-Propagation (TOP) detector is a key system for
charged-particle identification (PID). Although TOP operated successfully during
initial Belle~II data taking, increasing luminosity and beam-induced background
led to more frequent single-event upsets (SEUs) in the radiation-exposed Zynq
systems-on-chip (SoCs). Resulting lockups of the embedded processing systems
(PSs) interrupted data acquisition. To mitigate this limitation, waveform
feature extraction, a critical task of the on-detector Zynq PS, was migrated to the off-detector
PCIe40 readout PCs (ROPCs). The new architecture bypasses the PS in the event
data path and performs feature extraction outside the detector radiation
environment. This change removed SEU-induced PS lockups from the event data
path and enabled stable operation of the TOP front-end
electronics and the Belle~II data-acquisition system under increased luminosity
and background conditions. The typical TOP deadtime decreased from about 1\%
to a level consistent with zero after the final firmware patch, and stable
operation was demonstrated at a Level-1 (L1) trigger rate of 30~kHz with a
microchannel-plate photomultiplier-tube (MCP-PMT) hit rate of approximately
5~MHz per PMT. This paper describes the migrated architecture, its
deployment and validation, and its operational and PID performance.
\end{abstract}


\end{frontmatter}

\section{Introduction}

Belle~II is the detector built to study collisions at SuperKEKB, a
high-luminosity asymmetric-energy electron--positron collider at
KEK~\cite{Ohnishi2013,Abe2010,Adachi2018}.
The experiment collects large samples of $B$ mesons and performs precision
measurements of their decays to study charge-parity ($CP$) violation in the $B$-meson sector
and to search for phenomena beyond the Standard Model~\cite{Kou2019}. To obtain
a data set substantially larger than those of the previous $B$ factories, the
SuperKEKB/Belle~II program targets an integrated luminosity of
50~ab$^{-1}$ and an instantaneous luminosity of
$6\times10^{35}$~cm$^{-2}$s$^{-1}$~\cite{Abe2010,Ohnishi2013}. Operation at high luminosity produces
substantial beam-induced background and requires the Belle~II subdetectors and
their readout systems to sustain high detector occupancy and trigger rates.

The Time-of-Propagation (TOP) detector provides charged-hadron particle
identification (PID)
in the barrel region of Belle~II and is important for improving the sensitivity
of many physics measurements. It identifies charged particles using the
spatial and temporal distributions of Cherenkov photons propagating through
synthetic fused-silica (quartz) radiator bars. A detailed description of the
detector and its performance is given in Ref.~\cite{Atmacan_2025}. The TOP front-end electronics use
waveform-digitizer application-specific integrated circuits (ASICs) based on switched-capacitor arrays to sample and
digitize signals from 8192 channels of microchannel-plate photomultiplier tubes
(MCP-PMTs)~\cite{Kotchetkov2019TOP}. The digitized waveforms are collected by
front-end boardstacks. The Standard Control Read-Out Data (SCROD) board on each
boardstack uses a Zynq system-on-chip (SoC) to aggregate the waveform data. In
the original architecture, waveform feature
extraction was performed in software by the processing system (PS) of this
SoC. The PS extracted photon arrival-time and pulse-height information from the merged
waveforms before the resulting compact hit data were transferred to the
Belle~II data-acquisition (DAQ) system. This configuration is referred to as
PS-based mode.

This architecture met the TOP operational requirements during the initial Belle~II data-taking runs. As luminosity and beam-induced background increased,
however, single-event upsets (SEUs) and the resulting PS lockups increasingly
interrupted TOP data acquisition. The PS-based processing also limited readout
throughput at high trigger rate and occupancy. Continued luminosity growth
therefore required a more robust readout architecture.

The PCI Express-based PCIe40 backend had been available to TOP since 2021, but the PS-based
system still met the operating requirements at that time. Migration became
necessary only after higher luminosity and background increased the frequency
of PS-related interruptions and the required processing rate.


To minimize SEU-related PS lockups in the event data path, TOP waveform feature
extraction was migrated from the SCROD PS to the off-detector readout PCs
(ROPCs). The
migrated system uses PS-bypass mode to transport the event data and
performs feature extraction in software on the ROPCs. This paper describes the
original and migrated architectures, their end-to-end deployment and
validation, and the operational and PID performance of the
new scheme.

\section{Belle~II TOP Readout Overview}

The TOP front-end readout is organized into 64 boardstacks, with four
boardstacks serving each of the 16 TOP modules.\footnote{The channel count is
$8$ channels per ASIC $\times 4$ ASICs per Carrier Board $\times 4$ Carrier
Boards per boardstack $\times 4$ boardstacks per TOP module $\times 16$ TOP
modules, giving 8192 channels in total.} Each boardstack consists of four ASIC
Carrier Boards connected to one SCROD board.
A Carrier Board contains four eight-channel Ice Ray Sampler version X (IRSX)
waveform-digitizer ASICs and a Zynq Z-7030 SoC. The Carrier
SoC is used only for data control, including control of the IRSX devices and
transfer of the digitized waveforms. The SCROD is the merger board of the
boardstack. Its Zynq Z-7045 SoC aggregates the data streams from the four Carrier Boards,
corresponding to 128 waveform channels. The Zynq SoCs integrate programmable
logic (PL) and an ARM-based PS. A detailed description and
qualification of this front-end system is given in
Ref.~\cite{Kotchetkov2019TOP}.

\begin{figure}[htbp]
  \centering
  \begin{minipage}[c]{0.45\linewidth}
    \centering
    \includegraphics[width=\linewidth]{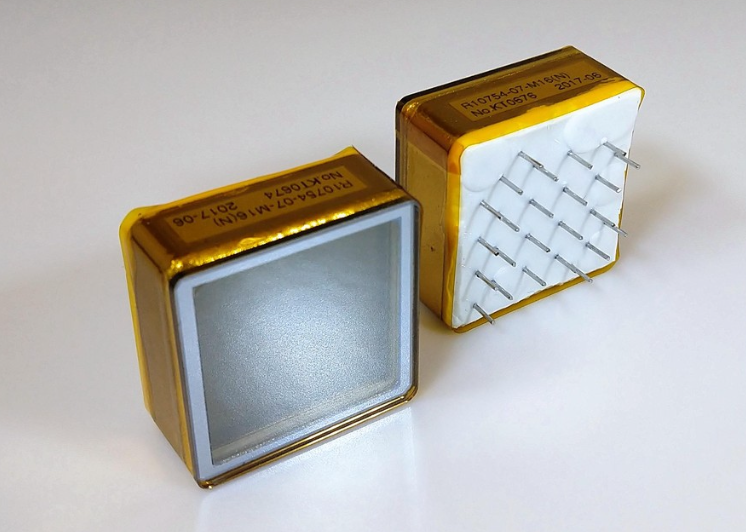}\\
    (a)
  \end{minipage}
  \hfill
  \begin{minipage}[c]{0.45\linewidth}
    \centering
    \includegraphics[width=\linewidth]{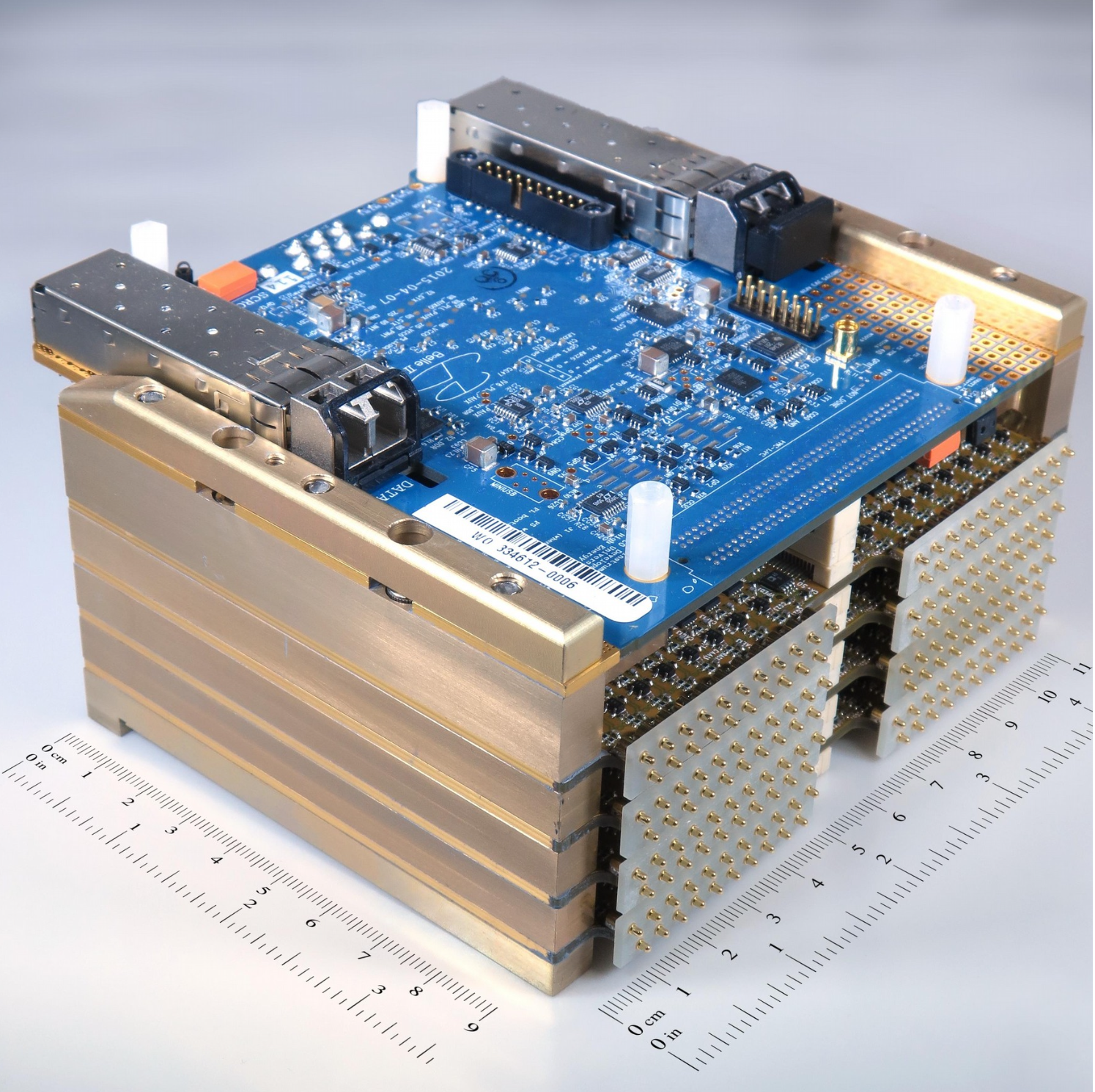}\\
    (b)
  \end{minipage}
  \caption{Photographs of (a) a 16-channel MCP-PMT used in TOP and
  (b) an assembled TOP front-end boardstack. One boardstack combines four
  Carrier Boards with one SCROD board and reads 128 MCP-PMT channels.}
  \label{fig:top_frontend_components}
\end{figure}

In this paper, a detector channel means one MCP-PMT signal channel and its
corresponding IRSX digitizer channel. A boardstack reads 128 detector channels.
Each channel of PCIe40 receives data from one boardstack via a single optical
input. Fig.~\ref{fig:top_frontend_components} shows an MCP-PMT and an assembled
TOP front-end boardstack.

The IRSX is a waveform-digitizer ASIC that uses a switched-capacitor array for
sampling at 2.714~GSa/s. Each channel contains 32,768 analog storage cells,
providing approximately 12~$\mu$s of sampling depth. The cells are addressed in
sampling windows of 64 consecutive samples. Each IRSX input has a comparator
with a threshold set by an on-chip digital-to-analog converter (DAC). When an
input signal exceeds this threshold, the comparator produces a channel-trigger
signal. The Carrier PL records the channel-trigger timing and correlates it
with the timing of a Level-1 (L1) trigger. For a matched channel trigger, the
Carrier PL identifies the corresponding region of interest in the analog
sample buffer and instructs the IRSX ASIC to digitize a fixed set of samples
around the candidate photon hit. In
this paper, such a selected sequence of digitized samples is called a waveform
segment; a segment can include samples from adjacent 64-sample windows. The
Carrier PL reads the selected waveform segments and transfers them to the SCROD over
boardstack-internal serial links. The IRSX readout is organized in channel
pairs, so the waveform from the partner channel is included with a selected
channel and can later be removed by feature-selection cuts.

\begin{figure}[htbp]
  \centering
  \includegraphics[width=\linewidth]{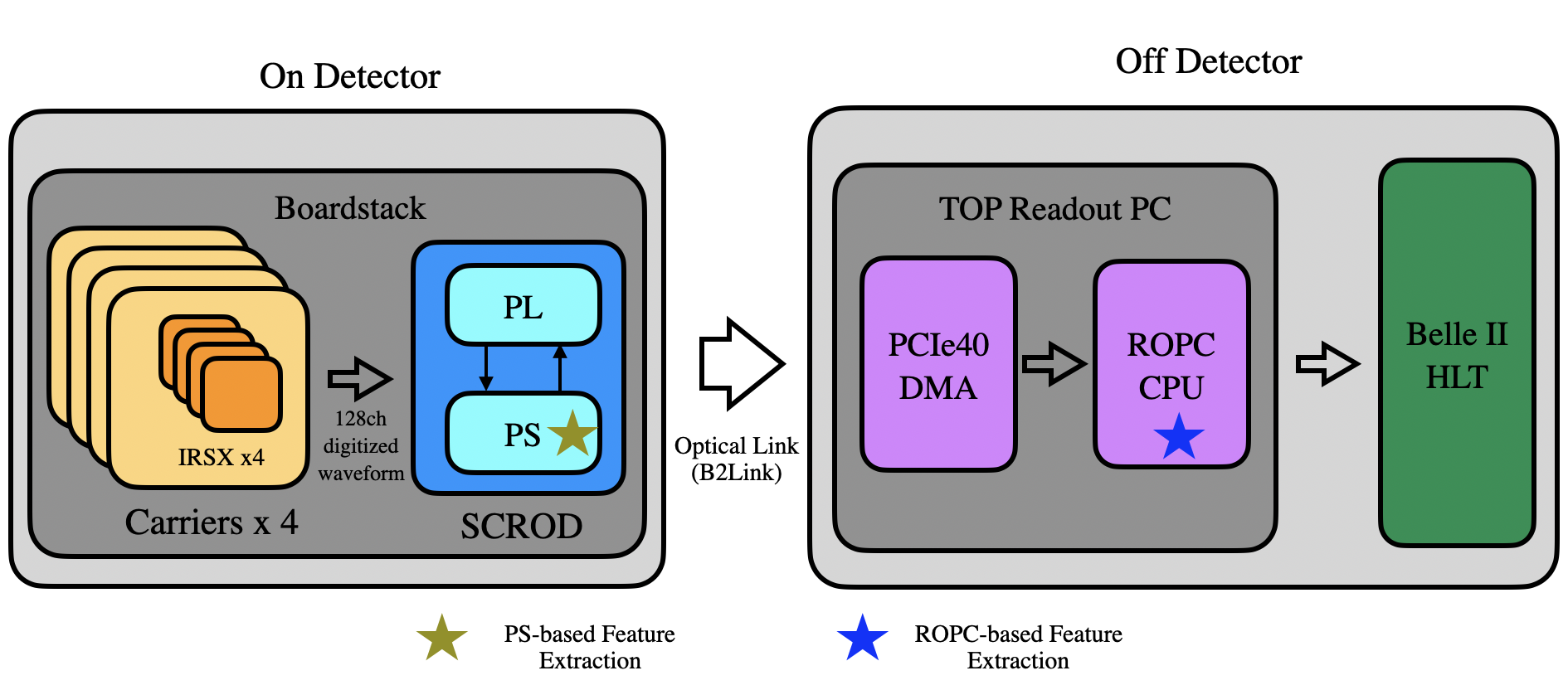}
  \caption{Data flow for one TOP front-end boardstack and the PCIe40-based
  backend. Four Carrier Boards digitize the selected waveforms, and the SCROD
  aggregates the data from 128 channels. In PS-based mode, the
  merged waveforms were passed from the SCROD PL to its PS for feature
  extraction. In PS-bypass mode, the waveform data are transmitted over the
  optical Belle2link to the PCIe40 board in the TOP ROPC, where feature
  extraction is performed on the ROPC central processing unit (CPU) before
  transfer to the Belle~II high-level trigger (HLT).}
  \label{fig:top_readout_data_flow}
\end{figure}

In PS-based mode, the SCROD PL collected the waveform segments for
an event into a packet containing the event and waveform metadata and
transferred it to the PS. After feature extraction, the compact
hit data were returned to the SCROD PL, which added the event header and
transmitted the packet via the optical Belle2link protocol. At the start of
Belle~II physics operation in 2019, these links were received by Common Pipelined
Platform for Electronics Readout (COPPER) modules~\cite{Yamada2017COPPER}.
Belle~II subsequently
developed and commissioned a PCIe40-based backend during 2020--2021, and TOP
began physics data taking with PCIe40 readout in autumn
2021~\cite{Zhou2021PCIe40,Lai2023PCIe40}. In the current configuration, the
SCROD optical links are received by PCIe40 boards installed in the TOP ROPCs.
TOP uses two ROPCs, each connected to 32 of the 64 boardstacks.
The PCIe40 boards use direct memory access (DMA) to transfer event data into
ROPC host memory. Fig.~\ref{fig:top_readout_data_flow} summarizes
the resulting end-to-end data path, with the original and migrated feature-extraction locations indicated.

In the original COPPER-based readout, the available backend throughput was
insufficient to transport digitized waveform segments at the required rate, so
feature extraction was performed in the SCROD PS before the data were
transmitted~\cite{Kotchetkov2019TOP}. Following the upgrade to PCIe40, its
higher aggregate throughput and DMA transfers to the ROPCs made it
possible to transport the digitized waveform segments and perform feature
extraction on the ROPCs~\cite{Zhou2021PCIe40,Lai2023PCIe40}. PS-bypass mode
does not transfer waveforms from all TOP channels; only waveform segments from
channels selected for each L1 trigger are transmitted. Nevertheless, sending
waveform segments instead of compact hit records increases the event payload.
This larger payload is the main bandwidth cost of moving feature extraction to
the ROPCs.


\FloatBarrier

\section{Feature Extraction in PS-Based Mode}

In PS-based mode, feature extraction was performed in software on the
SCROD PS. The PS unpacked each event packet and processed its selected
digitized waveform segments individually. For each segment,
the software applied sample-by-sample pedestal subtraction and used constant
fraction discrimination (CFD) to estimate the photon arrival time. Pulse-height
and pulse-shape information were derived at the same stage to form a compact
hit record~\cite{Gedcke1967CFD,Kotchetkov2019TOP}.

\begin{figure}[htbp]
  \centering
  \includegraphics[width=0.78\linewidth]{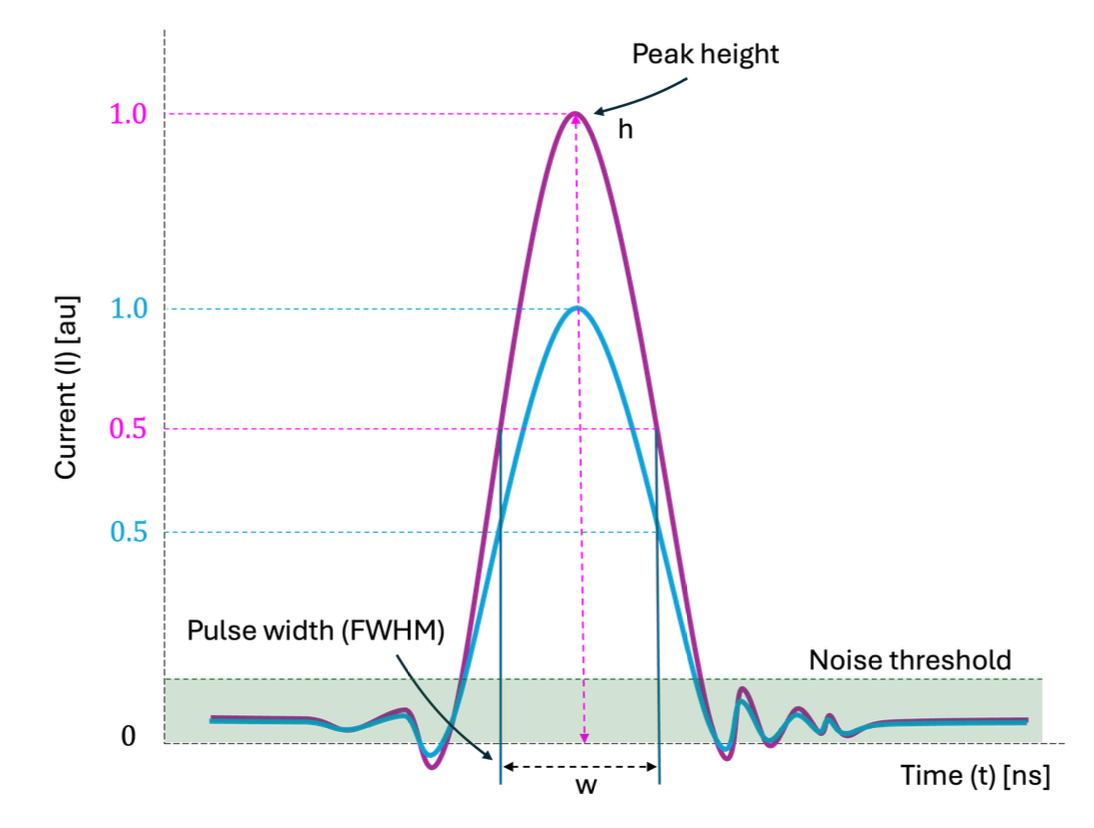}
  \caption{Schematic illustration of pulse observables used in waveform
  feature extraction. The peak height is denoted by $h$, and $w$ is the full
  width at half maximum. The colored 0.5 levels illustrate the constant
  fraction for pulses of different amplitudes, while the shaded region marks
  the noise-threshold range.}
  \label{fig:cfd_algorithm}
\end{figure}

The principal pulse observables are illustrated schematically in
Fig.~\ref{fig:cfd_algorithm}. The constant-fraction level is defined relative
to the height of each pulse; the implementation shown here uses a fraction of
0.5. The photon time is obtained from the crossing of
this level on the rising edge, reducing the amplitude dependence of the timing
estimate. The CFD algorithm does not determine the time of the pulse maximum;
it determines the leading-edge crossing time at a fixed fraction of the pulse
height. Its use does not assume identical pulse rise times: CFD reduces
amplitude-dependent time walk, but does not by itself remove timing shifts
caused by pulse-shape variations. Pulse width is therefore retained as a
separate observable for pulse selection.

Because the algorithm ran in software on the SCROD PS, the event
processing time increased with the number of waveform segments. At higher L1
trigger rates and detector occupancy, PS-based feature extraction therefore
became a throughput bottleneck and contributed to TOP deadtime.

The radiation-exposed PS also introduced a separate operational vulnerability.
Single-event upsets could cause a PS lockup and stop data flow from the affected
boardstack until recovery. The combination of processing-induced deadtime and
SEU-related interruptions motivated removal of the SCROD PS from the event data
path while preserving the established feature-extraction output.


\FloatBarrier

\section{ROPC-Based Feature-Extraction Architecture}

The migrated architecture keeps waveform acquisition on the detector and
moves feature extraction to the backend. In PS-bypass mode, the SCROD sends
the selected waveform segments directly to PCIe40. PCIe40 transfers them by
DMA to the ROPC host memory, where the TOP ROPC software produces the same
compact hit format used by the rest of the Belle~II DAQ. The complete path is
shown in Fig.~\ref{fig:top_readout_data_flow}.

\subsection{PS-bypass firmware}

\begin{figure}[htbp]
  \centering
  \includegraphics[width=\textwidth]{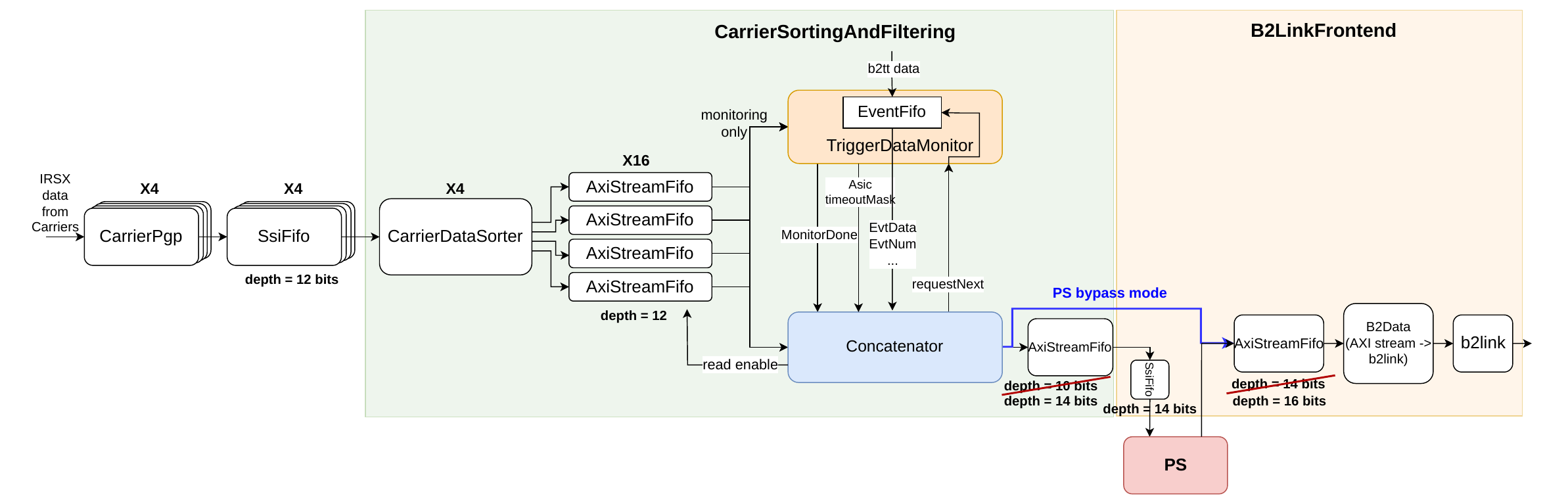}
  \caption{Event-data path in the SCROD firmware. Data from the four Carrier
  Boards are sorted into 16 IRSX streams and buffered before the Concatenator.
  In PS-based mode, the Concatenator output is sent through the PS input FIFO to the
  PS for feature extraction, and the resulting hit data return to the
  Belle2link frontend. In PS-bypass mode (blue path), the Concatenator output
  is sent directly to the Belle2link frontend. The crossed-out FIFO depths
  show values used before the PS-bypass buffering was increased.}
  \label{fig:scrod_dataflow}
\end{figure}

Fig.~\ref{fig:scrod_dataflow} shows the event-data path in the SCROD PL. The
CarrierDataSorter separates the incoming Carrier data into 16 streams, one for
each IRSX ASIC, and stores them in streaming first-in, first-out (FIFO)
buffers. After an L1 trigger, the
Concatenator collects the selected waveform data and forms one event packet.

\begin{figure}[htbp]
  \centering
  \includegraphics[width=\textwidth]{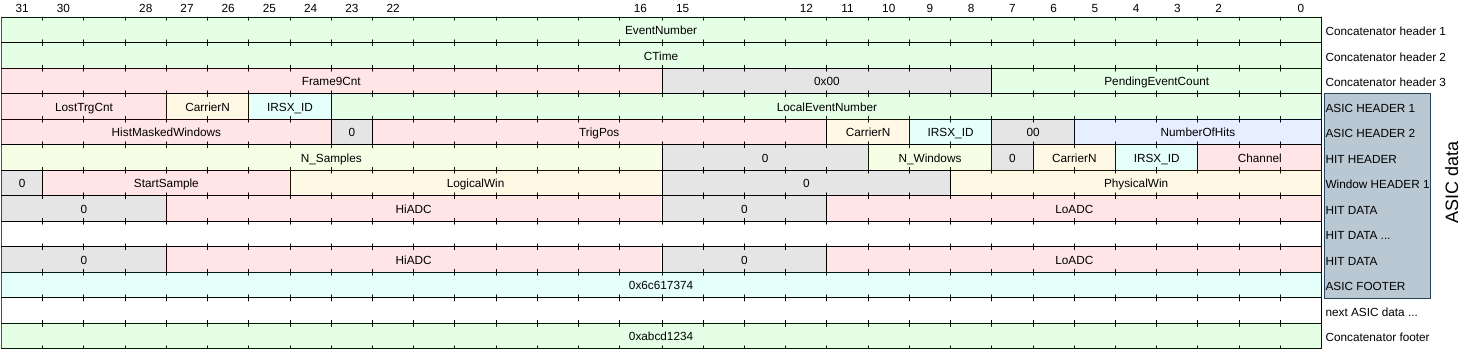}
  \caption{Format of the event packet produced by the Concatenator. The packet
  aggregates the selected waveform segments from one boardstack for one L1
  trigger. Three
  event-header words are followed by one block for each IRSX ASIC. Each ASIC
  block contains ASIC and hit headers, waveform-window information, digitized
  samples, and an ASIC footer. A Concatenator footer terminates the event. The
  bit positions of each 32-bit word are shown along the top.}
  \label{fig:concat_format}
\end{figure}

The packet produced by the Concatenator is shown in
Fig.~\ref{fig:concat_format}. In PS-based mode, this packet is sent to the
SCROD PS. The PS unpacks the packet, extracts pulse features from each waveform
segment, and returns compact hit records to
the PL at the input of the Belle2link frontend. In PS-bypass mode, the PL
routes the same Concatenator packet directly to the frontend, without using the
PS.

In PS-based mode, the PS memory provides intermediate event buffering, so the
PL input and output FIFOs can be relatively shallow. PS-bypass mode sends the
larger waveform packet directly to Belle2link and therefore requires more
buffering before transmission. The output FIFO depth was increased for this
purpose. As indicated in Fig.~\ref{fig:scrod_dataflow}, the address widths of
the two downstream FIFOs were increased from 10 to 14 bits and from 14 to
16 bits, respectively.

Early PS-bypass tests still found rare corrupted and desynchronized events at
high occupancy. In these events, the packet was larger than the free space in
the output FIFO. The firmware was changed to pause event processing until
enough FIFO space is available for the complete event. This backpressure
mechanism, together with the increased FIFO depth, removed the observed event
fragmentation.

\subsection{TOP ROPC software}

The TOP ROPC software performs the processing that was previously carried out
by the SCROD PS. It checks and unpacks each event, applies pedestal subtraction,
extracts pulse features with the CFD algorithm, applies pulse-height and
pulse-width cuts, and writes the standard compact TOP hit format. The output
format is unchanged, so no changes are required in downstream Belle~II event
processing.

TOP uses two ROPCs, each receiving data from 32 boardstacks. Each ROPC is
equipped with an Intel Xeon E5-2640 v4 processor with 10 physical cores and 20
logical CPUs, corresponding to two hardware threads per core. Each boardstack
stream is assigned to one PCIe40 input and one subevent builder;
thus, 32 subevent builders run on each ROPC. Detector mapping uses the
boardstack identifier in the event header rather than the PCIe40 channel
number.

\begin{figure}[htbp]
  \centering
  \includegraphics[width=\linewidth]{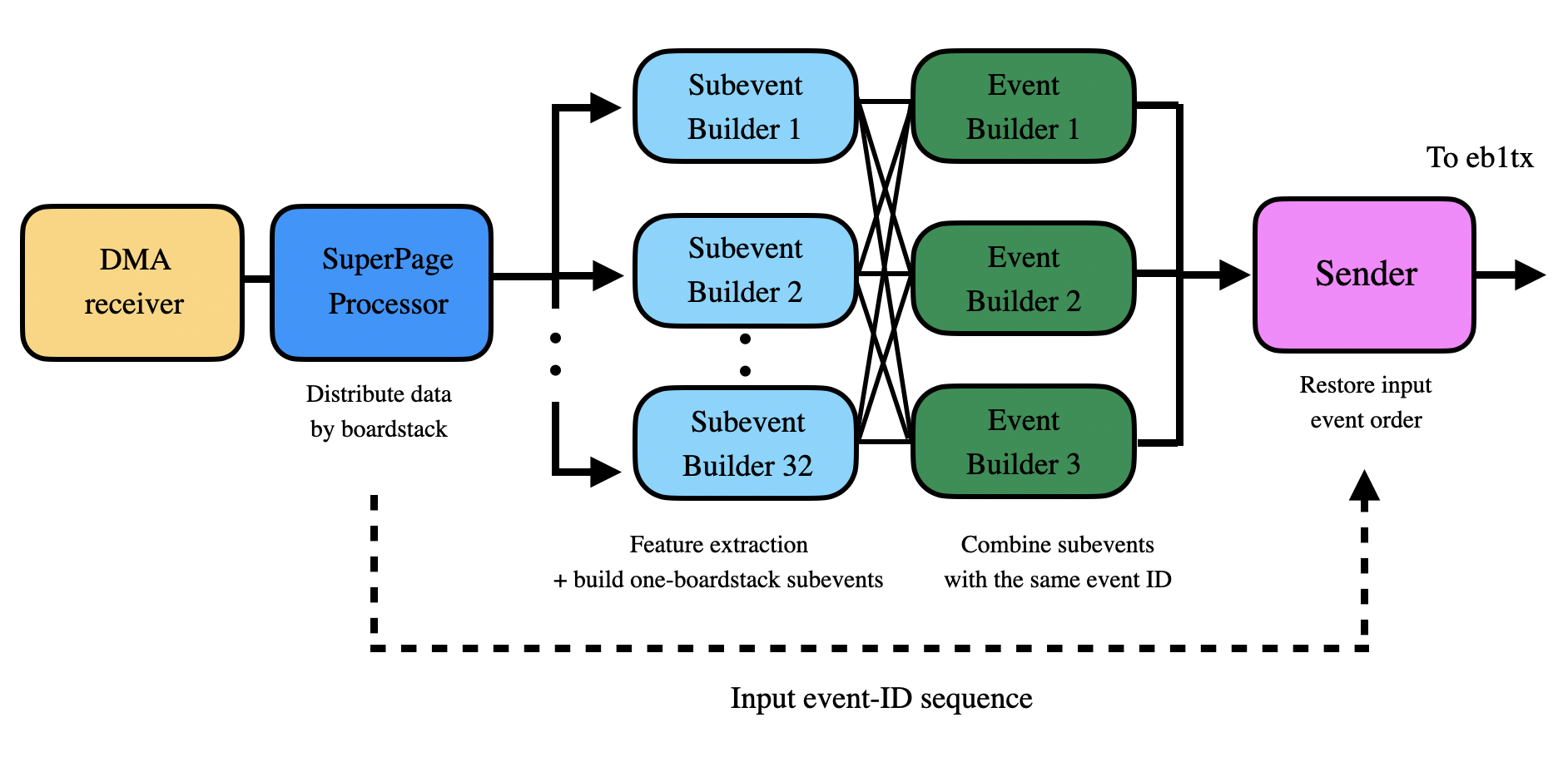}
  \caption{Parallel TOP processing in one ROPC. The SuperPage processor
  distributes data to 32 subevent builders, one for each boardstack connected
  to the ROPC. Each subevent builder performs pedestal subtraction and feature
  extraction and forms a one-boardstack subevent. Three event builders combine subevents with the same
  event ID. The dashed path carries the input event-ID sequence to the sender,
  which restores the original event order before transmission to
  \texttt{eb1tx}.}
  \label{fig:ropc_fe_module}
\end{figure}

The processing flow is shown in Fig.~\ref{fig:ropc_fe_module}. DMA places data
in large host-memory data buffers. The SuperPage processor distributes these
buffers among parallel subevent builders according to boardstack. Each
subevent builder uses the pedestal constants for its assigned boardstack to
perform sample-by-sample pedestal subtraction. It then applies the CFD and
pulse-selection algorithms and forms a subevent containing the resulting hit
data from that boardstack. An event identifier (event ID) associates subevents belonging
to the same event. Three event builders collect subevents with the same event
ID and combine them into complete TOP events. Because several events are processed in
parallel, they may be completed out of order. The sender uses the input event-ID
sequence recorded by the SuperPage processor to restore the original order and
transmits the events to the downstream Belle~II event-building process,
\texttt{eb1tx}, using the ZeroMQ messaging library~\cite{ZeroMQ}.

For backward compatibility, the module first checks the input data type. Data
that have already been processed in PS-based mode pass through unchanged. For
PS-bypass data, the module reads the raw waveform packet and the boardstack
identifier, performs feature extraction, and writes the same output format as
the PS-based system. Compatibility is maintained by using the same pedestal
values, CFD algorithm, pulse-height and pulse-width cuts, data-format checks,
and output packing.

PS-based processing remains available as a rollback mode because the ROPC
module recognizes compact PS-processed input and passes it through unchanged.
Malformed headers and checksum errors are detected during unpacking.


Pedestal constants are produced after each TOP power cycle and are copied to
the ROPCs. Each ROPC stores constants for its 32 boardstacks; the pedestal data
occupy approximately 500~MB per ROPC. When the ROPC software starts, the
constants are loaded in parallel and indexed by boardstack identifier. Loading
takes less than 10~s. The measured pedestal values have remained stable for
several weeks without a significant change in feature-extraction performance;
nevertheless, the operating procedure conservatively refreshes the files every
few days.

After pedestal subtraction, the waveform samples are passed to the CFD module.
The Carrier Boards transmit waveform data in channel pairs. Pulse-height and
pulse-width cuts suppress noise and remove records for an empty partner
channel. The accepted features are then packed into the standard TOP data
format and returned to the ROPC event-building process.

Moving feature extraction to the ROPCs removes waveform processing from the
radiation-exposed SCROD PS. It also allows the algorithm to be updated and
monitored in the backend software environment.

\FloatBarrier

\section{Deployment and Validation}

Before deployment, data recorded in PS-bypass mode were compared with data
recorded in PS-based mode. The tests checked pulse observables, pedestal
constants, channel masks, and event integrity. They identified inconsistent
channel-mask information in the pedestal files, which was corrected before
deployment. Without this correction, the PS and ROPC paths could have applied
different active-channel selections to the same boardstack. Event-format and
synchronization checks were also used to identify malformed packets and
event-ID mismatches among boardstacks.

\begin{figure}[htbp]
  \centering
  \includegraphics[width=0.96\linewidth]{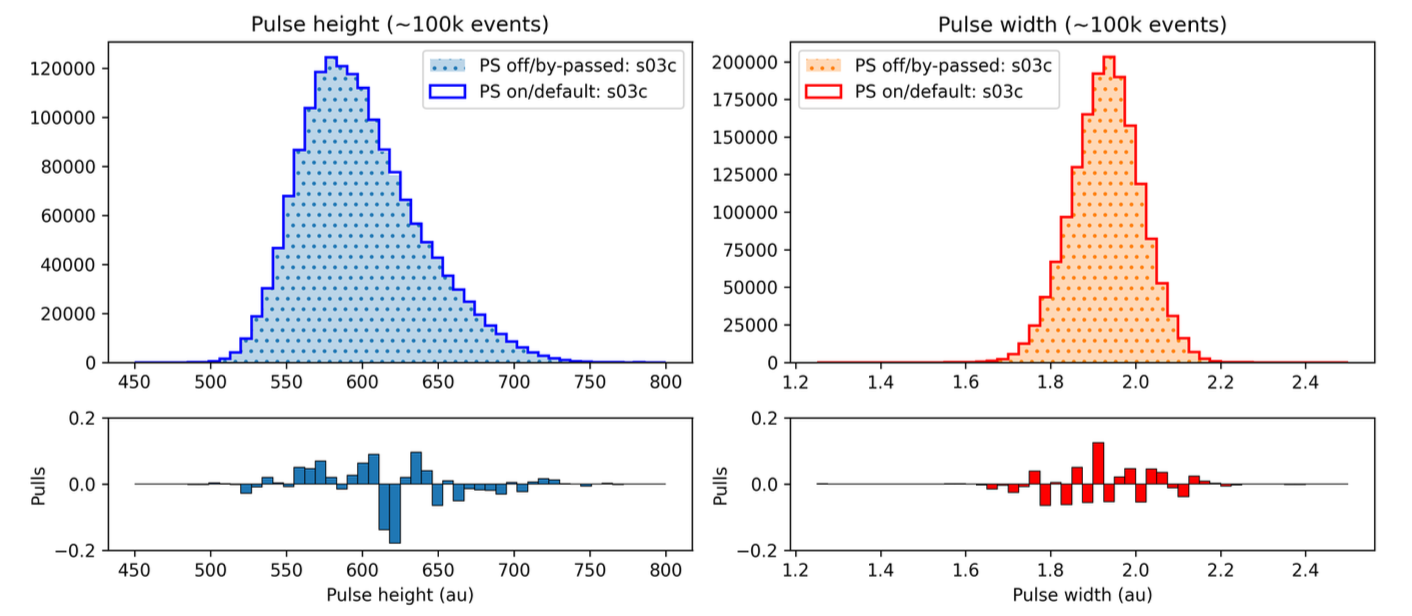}
  \caption{Comparison of the pulse-height (left) and pulse-width (right)
  distributions for approximately $10^5$ events from boardstack s03c with the
  readout operating in PS-based mode and PS-bypass mode. The lower panels show
  the bin-by-bin comparison quantity labeled ``Pulls'' in the original
  validation plot; values near zero indicate agreement between the two
  distributions. The plot labels ``PS on/default'' and ``PS off/by-passed''
  correspond to PS-based mode and PS-bypass mode, respectively. The
  abbreviation a.u. denotes arbitrary units.}
  \label{fig:ps_on_off_comparison}
\end{figure}

Fig.~\ref{fig:ps_on_off_comparison} shows a representative comparison of
pulse-height and pulse-width distributions obtained in PS-based mode and
PS-bypass mode.

Tests at high trigger rate and detector occupancy revealed occasional corrupted
events. This observation led to the larger FIFO and backpressure changes
described in Section~4.1. The end-to-end tests were then repeated under the same
high-load conditions to verify event integrity before deployment.


\FloatBarrier

\section{Operational Performance}

\subsection{Performance during physics data taking}

\begin{figure}[htbp]
  \centering
  \includegraphics[width=0.98\linewidth]{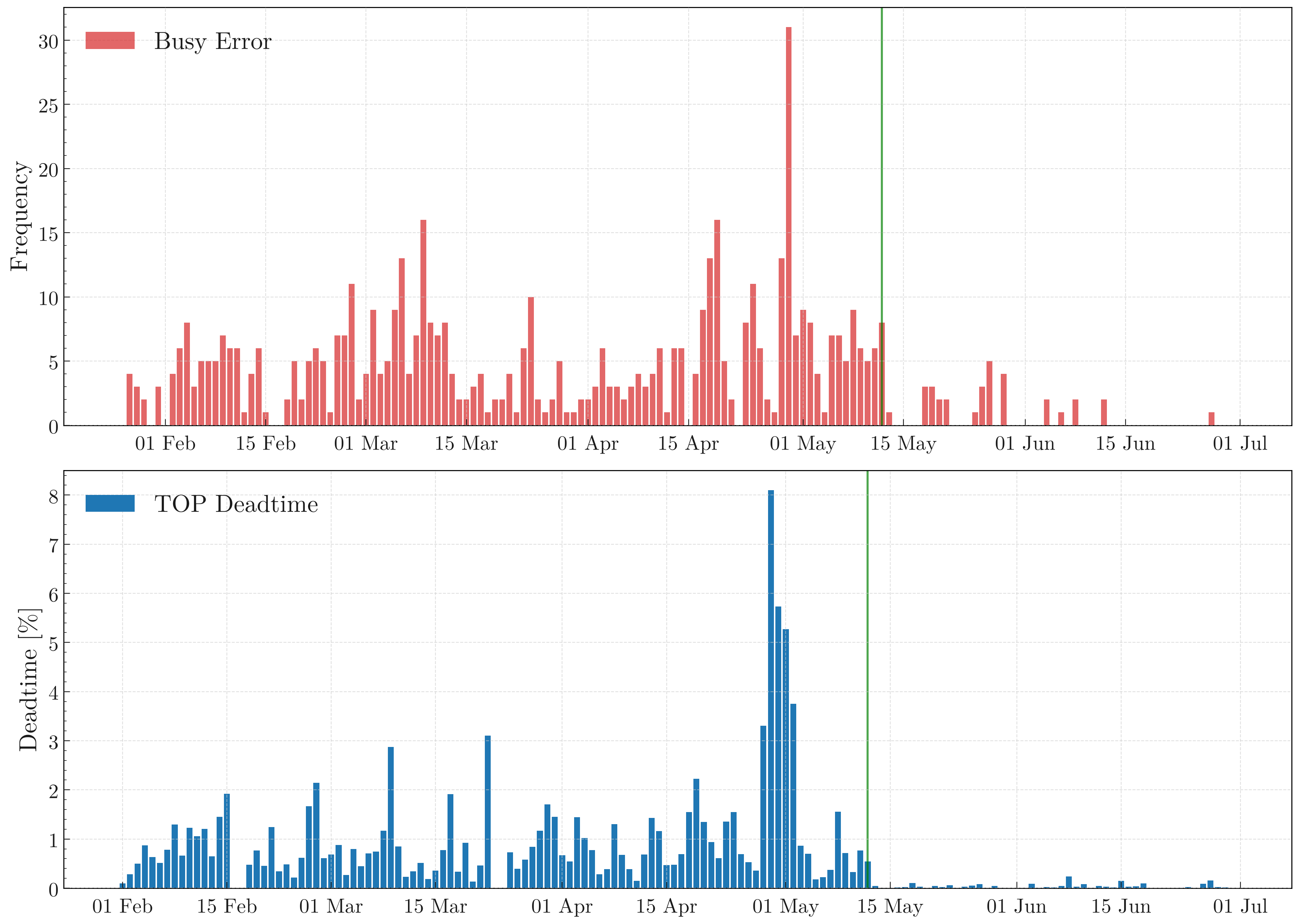}
  \caption{TOP operational performance from February to early July 2026. The upper
  panel shows the number of run stops associated with TOP-busy errors in each
  plotted time bin, and the
  lower panel shows the TOP deadtime. TOP operated in PS-based mode before
  May 12 and in PS-bypass mode from May 12. The green vertical line marks this
  transition. A firmware patch was applied on May 30; this date is not marked
  separately in the figure.}
  \label{fig:operational_performance}
\end{figure}

Following deployment, TOP operated with feature extraction on the ROPCs during
Belle~II data taking. Removing PS-based feature extraction from the event path
substantially reduced the deadtime associated with front-end processing. This
allowed stable TOP operation with the standard higher-rate trigger
configuration. Fig.~\ref{fig:operational_performance} shows the TOP-related run
stops and deadtime before and after the migration.

Most TOP-busy errors during operation in PS-based mode were caused by
SEU-induced lockups of the SCROD PS. PS-bypass mode removed this failure
mechanism by excluding the PS from the event data path. A smaller number of
TOP-busy errors remained during the initial PS-bypass operation and became
sparse after the firmware patch. The TOP deadtime, which was typically at the
1\% level in PS-based mode, became consistent with zero after the patch.
Deadtime excursions exceeding 5\% were observed in PS-based mode but not in
PS-bypass mode.

The accelerator and background conditions changed during the period shown in
Fig.~\ref{fig:operational_performance}. The run-stop counts are not normalized
to live time or integrated luminosity, so the figure is used as an operational
time history rather than as a measurement of an absolute failure-rate ratio.
The connection to the migration is supported by the error diagnosis: the
dominant pre-migration TOP-busy errors were traced to SCROD PS lockups, a path
that is not used in PS-bypass mode.

Under the current physics-running conditions, with an L1 trigger rate of about
12~kHz and a typical MCP-PMT hit rate of about 5~MHz per PMT, the TOP readout
process uses at most approximately 550\% CPU in the operating-system
convention, equivalent to 5.5 fully utilized logical CPUs.


\FloatBarrier

\subsection{High-rate performance}
\label{sec:high_rate_performance}

A dedicated high-rate study relaxed the electromagnetic calorimeter (ECL)
trigger requirements to raise the L1 rate while the MCP-PMT hit rate was held
at approximately 5~MHz per PMT. Stable TOP operation was confirmed up to an L1
trigger rate of 30~kHz without a significant increase in TOP deadtime or
readout instability. At this point, the TOP readout process used approximately
1000\% CPU, equivalent to 10 fully utilized logical CPUs. The test therefore shows that
the migrated readout can sustain the Belle~II design L1 rate at the tested
background level. Operation at the same trigger rate with the higher detector
occupancy expected at larger luminosity will require additional processing
margin, as discussed in Section~\ref{sec:future_developments}.


\section{Particle-Identification Performance}

The feature-extraction migration must preserve the timing and pulse-height
information used by the TOP reconstruction and PID algorithms.

The comparison uses relatively small data samples of kinematically tagged
kaons and pions from $D^0 \to K^-\pi^+$ decays. To suppress combinatorial
background, the $D^0$ candidates are reconstructed through the decay
$D^{*+} \to D^0\pi^+$, and the $D^{*+}$ candidates are required to have a
momentum greater than 2.5~GeV/$c$ in the center-of-mass frame.

\begin{figure}[htbp]
  \centering
  \includegraphics[width=0.62\linewidth]{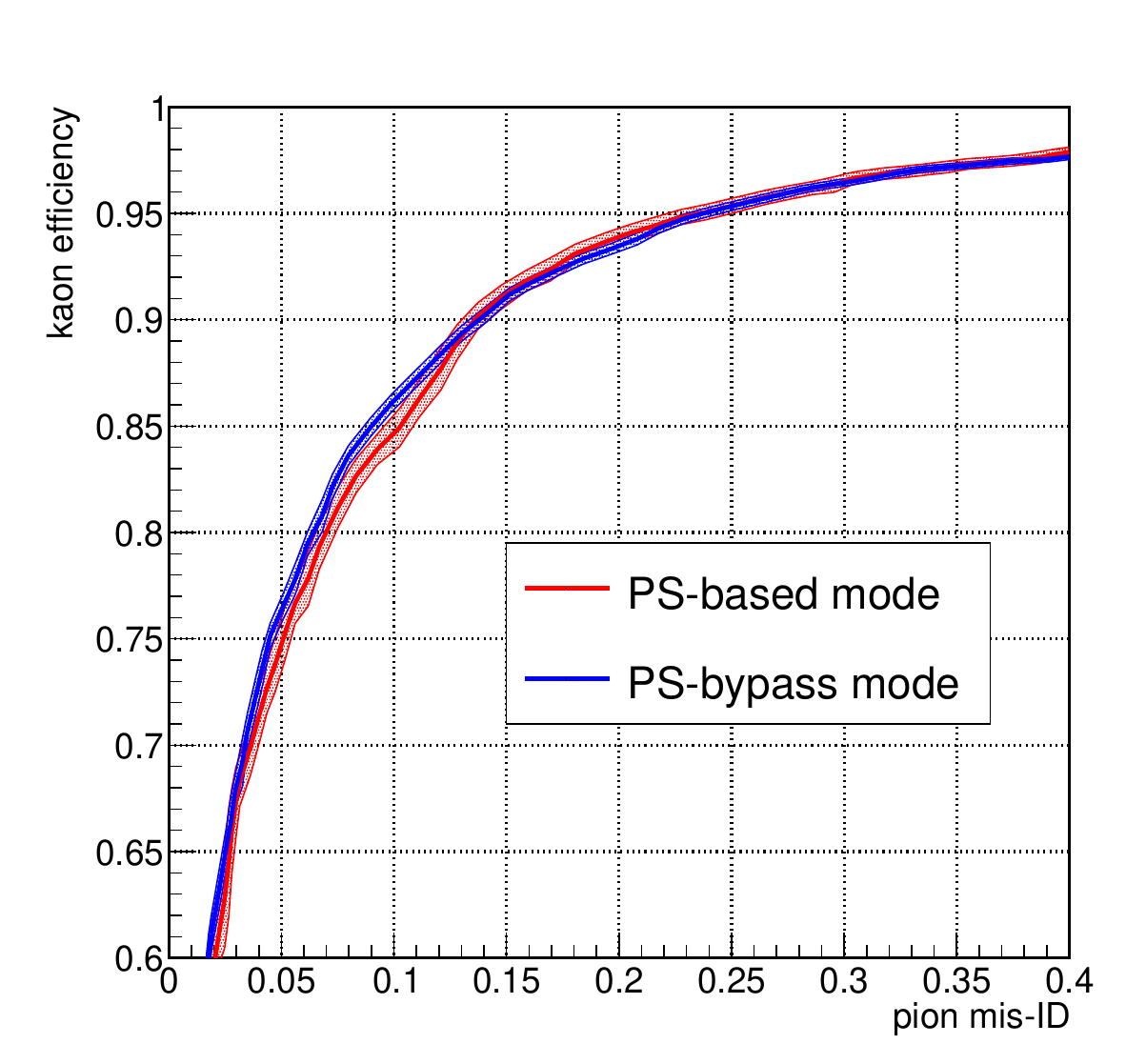}
  \caption{Kaon-identification efficiency as a function of the pion
  misidentification probability obtained from Belle~II physics data in
  PS-based mode (red) and PS-bypass mode (blue). The samples contain
  approximately 3400 and 10\,000 selected $D^0$ decays, respectively. The
  curves are obtained by scanning the binary-PID selection threshold, and the
  bands indicate the $\pm1\sigma$ statistical uncertainties.}
  \label{fig:pid_performance}
\end{figure}

Fig.~\ref{fig:pid_performance} compares the TOP kaon-identification
efficiency as a function of the pion misidentification probability using
Belle~II physics data collected before the migration in PS-based mode and after
the migration in PS-bypass mode. These receiver-operating-characteristic (ROC)
curves are obtained by scanning the selection threshold of the binary PID
discriminator; each position along a curve therefore represents a different
threshold.

The curves obtained in the two modes are consistent with each other
within the statistical fluctuations associated with the limited data sample
sizes. This demonstrates consistent TOP PID performance before and after the feature-extraction
migration.


\FloatBarrier

\section{Future Developments}
\label{sec:future_developments}

The ROPC processing load depends on both the L1 trigger rate and the detector
occupancy. The L1 rate sets the number of events processed per second, while
the MCP-PMT hit rate affects the number and complexity of the waveform segments
in each event. As discussed in Section~\ref{sec:high_rate_performance}, the
high-rate study reached 30~kHz at
approximately 5~MHz per PMT while the TOP readout process used the equivalent
of about 10 fully utilized logical CPUs. Future
high-luminosity operation may require operation at
the design L1 trigger rate of 30~kHz, or potentially higher rates, together with
PMT hit rates of approximately 10~MHz per PMT or higher. The present
implementation may therefore have insufficient processing margin under those
combined conditions.

Additional CPU or graphics processing unit (GPU) resources and software
optimization are being studied as near-term improvements.
The CFD algorithm must also handle pileup
and beam-background noise, both of which become more common as occupancy
increases.

Artificial intelligence and machine learning (AI/ML) methods are being explored for
longer-term feature extraction on the ROPCs. One approach is a lightweight model that directly extracts photon
timing and pulse height. Another is a classifier that identifies noise and
multi-pulse waveforms and then selects suitable inputs or parameters for the
existing feature-extraction algorithm.

The first step is to prepare representative waveform samples containing
pileup and beam-background conditions for offline training and validation.
Candidate models will be compared with the current CFD algorithm in timing
resolution, particle-identification performance, and CPU use. A selected model
would then be optimized for ROPC inference and run in parallel with CFD for
end-to-end validation before deployment. No AI/ML-based performance results
are reported here.

\section{Summary}

Feature extraction for the Belle~II TOP detector has been migrated from the
Zynq processing system on the SCROD boards to the PCIe40 readout PCs. The new
architecture operates in PS-bypass mode while retaining CFD-based extraction
of photon timing and pulse-height information on the ROPCs. It removes an
operationally vulnerable processing stage from the radiation-exposed front end
and provides a more flexible backend environment for waveform processing.

End-to-end deployment tests identified problems in pedestal and masking
handling and, under physics-like high-rate and high-occupancy conditions, rare
data corruption in the PS-bypass data path. Increasing the event FIFO capacity
and adding backpressure resolved the observed problem. Operation in PS-bypass
mode substantially improved TOP
readout stability, and comparisons with PS-based mode show consistent PID
performance.

The migration demonstrates the importance of validating the full readout and
configuration chain under conditions representative of physics data taking. The
ROPC processing platform also provides a basis for future development of
more advanced waveform-processing algorithms.


\section*{Acknowledgements}

The authors thank the SuperKEKB team for delivering high-luminosity
collisions, the Belle~II operation team for stable detector operation, and the
Belle~II data-acquisition team for the development and operation of the
PCIe40-based readout system.

The construction and operation of Belle~II are supported in part by the
Ministry of Education, Culture, Sports, Science and Technology (MEXT) of
Japan. This work was supported in part by the Slovenian Research Agency under
Grants No.~J1-50010 and No.~P1-0135, and by the U.S. Department of Energy,
Office of High Energy Physics, under Awards No.~DE-SC0010504,
No.~DE-SC0011784, and No.~DE-SC0007914.

\bibliographystyle{elsarticle-num}
\bibliography{references}

\end{document}